\documentclass[12pt,a4paper,british,superscriptaddress]{iopart}
\usepackage[T1]{fontenc}
\usepackage[latin9]{inputenc}
\usepackage{babel}
\usepackage{graphicx}
\usepackage{epstopdf}
\usepackage[numbers,square,sort&compress]{natbib}
\usepackage[unicode=true,pdfusetitle,bookmarks=false,breaklinks=false,pdfborder={0 0 1},backref=false,colorlinks=false]{hyperref}
\usepackage{array}
\usepackage{verbatim}
\usepackage{amsfonts}
\usepackage{amssymb}
\usepackage{color}
\usepackage{iopams}
\usepackage{bibentry}
\usepackage{revsymb}

\newcommand{\bra}[1]{\langle #1|}
\newcommand{\ket}[1]{| #1 \rangle}
\newcommand{\braket}[2]{\langle #1 | #2 \rangle }

\renewcommand{\Tr}{\mathrm{Tr}}
\renewcommand{\t}[1]{\mathrm{#1}}

\renewcommand{\comment}[1]{}

\begin{document}
\title{Bayesian quantum frequency estimation in presence of collective dephasing}
\author{Katarzyna Macieszczak$^{1,2}$, Martin Fraas$^3$, Rafa{\l} Demkowicz-Dobrza{\'n}ski$^2$}
\address{$^1$ University of Nottingham, School of Mathematical Sciences, University Park, NG7 2RD Nottingham, UK}
\address{$^2$ Faculty of Physics, University of Warsaw, ul. Ho\.{z}a 69, PL-00-681 Warszawa, Poland}
\address{$^3$ Theoretische Physik, ETH Z{\"u}rich, 8093 Z{\"u}rich, Switzerland}
\nobibliography*
\begin{abstract}
We advocate a Bayesian approach to optimal quantum frequency estimation - an important issue for future quantum enhanced atomic clock operation.
The approach provides a clear insight into the interplay between decoherence and the extent of the prior knowledge
in determining the optimal interrogation times and optimal estimation strategies.
We propose a general framework capable of describing local oscillator noise as well as additional collective atomic dephasing effects.
For a Gaussian noise the average Bayesian cost can be expressed using the quantum Fisher information
and thus we establish a direct link between the two, often competing, approaches to quantum estimation theory.
\end{abstract}
\pacs{03.65.Ta, 06.30.Ft}

\maketitle
\section{Introduction}
\begin{figure}[t]
\begin{flushright}
\includegraphics[width=0.8\columnwidth]{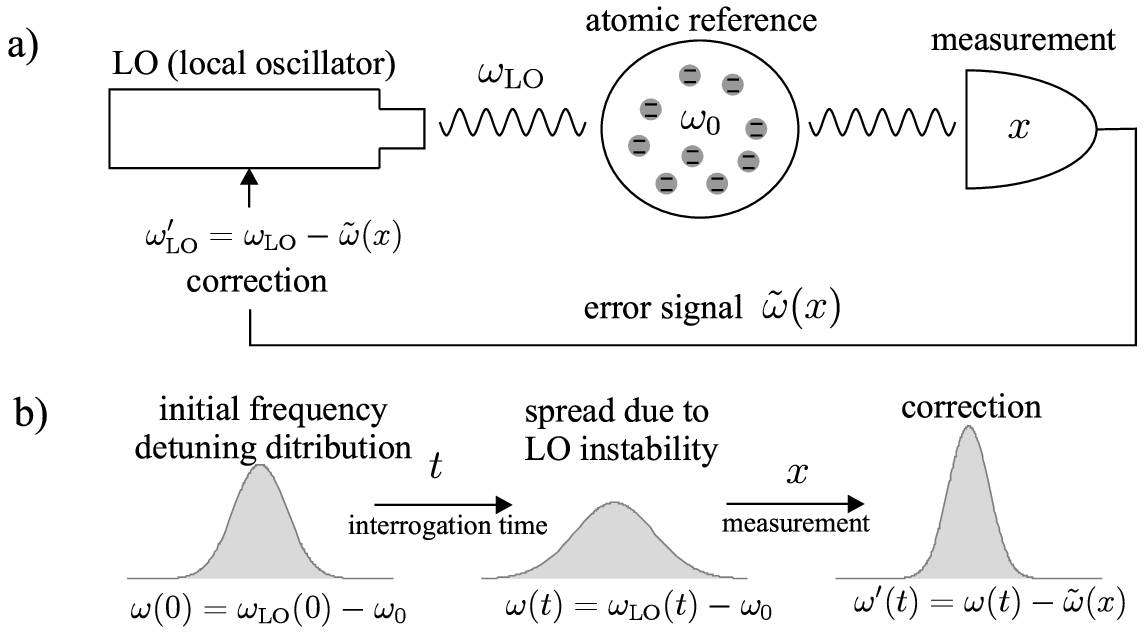}
\caption{a) Basic scheme of atomic clock operation. LO coupled to an atomic reference generates an error
signal on frequency difference $\omega = \omega_{\t{LO}}-\omega_0$ which is used to keep the LO locked to the atomic transition.
b) Prior distribution of $\omega$ spreads while atoms are being interrogated due to LO instability.
Estimation and a feedback correction procedure help to keep the $\omega$ distribution narrow. }
\label{fig:scheme}
\end{flushright}
\end{figure}

 Modern atomic clocks allow for time keeping with instability better
than $10^{-15}$, and are reaching towards new applications in geodesy \cite{Kleppner2006} and tests of fundamental physics \cite{Chou2010}. As of 2013 the best atomic clock have instability $10^{-18}$ after 7 hours of averaging \cite{Hinkley2013}.
Crystal oscillators or stabilized lasers have excellent short-time
frequency stability but their frequency $\omega_{\t{LO}}(t)$ tends to drift due to variations of temperature or stress and it needs
to be locked to a narrow atomic transition $\omega_0$, between two atomic levels
$\ket{0}$, $\ket{1}$, in order to guarantee long-time stability, see Fig.~\ref{fig:scheme}.  The resulting stability is limited by local oscillator (LO) noise and the atomic signal to noise ratio. A single estimation strategy have to be used periodically to determine the frequency offset $\omega(t) = \omega_{\t{LO}}(t) - \omega_0$. Knowledge of the LO noise and the value of $\omega$ from the previous feedback cycle provides a natural prior, making Bayesian analysis well suited for a study of the optimal estimation protocol.

In a typical Ramsey interferometric scheme $N$ atoms interact with an external electromagnetic field
evolving at frequency $\omega$, which for now is a time independent parameter. First, the atoms experience a $\pi/2$-pulse which transforms the ground state $\ket{0}$ of each one into
the $(\ket{0}+\ket{1})/\sqrt{2}$ superposition. After evolving freely for a time $t$, they experience another $\pi/2$-pulse, subsequently being subjected to a measurement determining the number $n_1$ of atoms
which made a transition to the excited state $\ket{1}$. In the case of perfect synchronization, $\omega_{\t{LO}}=\omega_0$,
all atoms end up in the $\ket{1}$ state, while in the presence of detuning
the average ratio reads $\langle n_1 \rangle /N = \cos^2[\omega t/2]$ \cite{Ramsey1980}.
The scheme is identical regardless of whether the $\omega_0$ transition is microwave (Cs-fountain clocks) or optical
(ion traps, optical lattices).
The procedure estimates the frequency difference $\omega$ on the basis of the number of excited atoms detected and performs a feedback on $\omega_{\t{LO}}$,
 locking it  to the atomic frequency $\omega_{0}$. Mathematically, the scheme is equivalent to
an $N$-photon  Mach-Zehnder interferometric experiment with a relative optical phase delay $\varphi = \omega t$ between the arms, which play the role of the two atomic levels \cite{Lee2002}. Just as
in the optical interferometry, fluctuation in the number of detected atoms gives rise to
the shot noise $1/\sqrt{N}$ scaling of the frequency estimation precision.
Shot noise scaling is a direct consequence of lack of correlations among the atoms.
If the atoms were prepared in a correlated quantum state, such as a spin-squeezed or a GHZ state,
   the $1/\sqrt{N}$ limit could be beaten and the $1/N$ Heisenberg bound approached, at least in idealized decoherence free scenarios
   \cite{Wineland1992, Holland1993, Bollinger1996, Berry2000, Lee2002, Giovannetti2006}.
While quantum enhanced sensing ideas have proved useful in a number of practical applications \cite{Roos2006, LIGO2011, Ospelkaus2011, Sewell2012},
it has been observed that in the presence of losses \cite{Caves1981, Huver2008, Dorner2009, Demkowicz2009, Knysh2010, Demkowicz2013}, dephasing
\cite{Huelga1997, Genoni2011, Dorner2012} or other decoherence processes \cite{Shaji2007, Sarovar2006} the maximum achievable quantum enhacement
is limited.
The ultimate goal of this type of research is to
study the performance of \emph{optimal} strategies in which probe states, measurements and inferring strategies are optimized
in a given sensing protocol \cite{Giovannetti2011, Banaszek2009}.
 The resulting optimal sensing precision may be then regarded as fundamental, i.e.
imposed by the laws of nature. Powerful tools are available
for the effective identification of these limits \cite{Escher2011, Demkowicz2012, Kolodynski2013}.
 A number of papers have studied the effectiveness of quantum strategies
in frequency estimation \cite{Huelga1997, Dorner2012, Chin2012, Szankowski2012} or, more specifically,
in atomic clock performance \cite{Buzek1999, Andre2004, Leibfried2004, Mullan2012, Borregaard2013b}.
These approaches, however, lacked generality. Some ignored the role of prior frequency distribution or the presence of decoherence. Others
studied a particular estimation scheme.
The aim of this paper is to
 propose a simple, general and effective Bayesian approach to the study  of general quantum enhancement schemes in frequency estimation. We show that finding the optimal estimation scheme in our setting has equivalent complexity as optimizing
quantum Fisher Information (QFI) \cite{Helstrom1976,Braunstein1994} and in the case of Gaussian priors we establish a direct link between these two approaches.
We also argue that the Bayesian approach is more fundamental
in the study of the  limitations on the performance of quantum enhanced atomic clocks - a point of view shared by other authors
\cite{Buzek1999,Mullan2012, Fraas2013, Mullan2014}.

\section{Cram{\'e}r-Rao bound approach}
In the case of a probe state $\rho_\omega$ with an encoded parameter $\omega$ to be estimated, the standard quantum Cram{\'e}r-Rao bound  (CRB) states
that, regardless of the measurements and unbiased estimators used, the estimation variance $\Delta^2 \widetilde{\omega}$  is lower bounded by
\begin{equation}
\label{eq:cr}
\Delta^2 \widetilde{\omega} \geq \frac{1}{F(\rho_\omega)}, \ F(\rho_\omega)=\Tr(\rho_{\omega} L_\omega^2),
 \ \frac{1}{2}\{L_\omega,\rho_{\omega}\} = \frac{\t{d} \rho_{\omega}}{\t{d} \omega},
\end{equation}
where $F$ is the QFI, $\{\,,\}$ is the anticommutator and $L_\omega$, implicitly defined above, is
 the symmetric logarithmic derivative of $\rho_\omega$.
When estimating frequency the probe state will typically be:
\begin{equation}
\label{eq:evolution}
\rho_\omega= e^{- \mathrm{i} H \omega t} \Lambda_t(\rho) e^{\mathrm{i} H \omega t},
\end{equation}
where $\rho$ is the interferometer input state, $t$ is the interrogation time, $H$ is the generator of the unitary evolution encoding
$\omega$, while $\Lambda_t$ represents decoherence processes. In this case, since $\frac{\t{d} \rho_{\omega}}{\t{d} \omega}=\mathrm{i} t [\rho_\omega, H]$,
$F(\rho_\omega)$ does not depend on $\omega$ and reads:
\begin{equation}
\label{eq:fisherH}
F(\rho_\omega)= F(\Lambda_t(\rho),H t)= 2t^2 \sum_{ij}\frac{|\bra{i}H\ket{j}|^2(\lambda_i - \lambda_j)^2}{\lambda_i+\lambda_j},
\end{equation}
where $\lambda_i$, $\ket{i}$ are the eigenvalues and the eigenvectors of $\Lambda_t(\rho)$ respectively.
The optimal strategy is obtained by maximizing $F$ leading to the optimal $\rho$, $t$ and the optimal projective measurement given by the
eigenbasis of $L_\omega$.
Although applied fruitfully in many
realistic metrological scenarios, including lossy interferometry \cite{Dorner2009, Knysh2010, Escher2011, Demkowicz2013} and
noisy frequency estimation \cite{Escher2011, Dorner2012, Szankowski2012, Chin2012}, CRB approach suffers from a number of drawbacks.
CRB is \emph{not saturable} in general.
It is only guaranteed to be saturable in special cases including Gaussian models, or in a repeated independent experiment framework (for a large number $k$ of experiments it is possible to achieve $\Delta^2 \widetilde{\omega} \approx \frac{1}{k F}$) \cite{Holevo1982}.
Moreover, since QFI is a \emph{local} quantity---for a given $\omega$ it only depends on $\rho_\omega$ and its first derivative---
it completely ignores any possible ambiguities in a reconstruction the frequency value from a phase value that may arise for a sufficiently broad prior parameter distribution.

In frequency estimation
the interrogation time $t$ is a controllable parameter subject to optimization.
 If $t$ is large,
even a very narrow prior distribution in $\omega$ may result in a phase $\varphi=\omega t$ distribution outside a local regime, or even broad enough
on $[0, 2\pi]$ interval for reconstruction ambiguities to become relevant.
Validity of the local regime, where CRB based conclusions hold, cannot be a priory assumed. It depends on all the details of estimation scheme, the LO noise, the interrogation time and the initial state.
 In contrast, the Bayesian approach allows for
full control of all the issues raised above, and any optimal scheme derived in these framework can be used as a universal benchmark.
In certain parameter regimes the Bayesian approach may produce results compatible with the QFI approach but this is not the case in general.
The Bayesian
approach has been present in quantum estimation literature from the beginning \cite{Helstrom1976}.
Nevertheless, strict Bayesian approach is often computationally challenging and rigorous
solutions are scarce and limited to decoherence-free scenarios \cite{Berry2000, Demkowicz2011}.
In this paper we show that the frequency estimation problem described above is efficiently solvable within the Bayesian framework even in the presence of arbitrarily time-correated collective dephasing  processes.

\section{Bayesian approach}
Let $p_\omega$ be the prior distribution and $\rho_\omega$ the evolved probe state. Without loss of generality we
assume  $\int \t{d} \omega \, p_\omega \omega $=0. The state
$\rho_\omega$ is subject to a POVM measurement \cite{Holevo1982} $\{\Pi_x\}$, $\Pi_x \geq 0$, $\int \t{d}x \Pi_x = \openone$
and the parameter $\omega$ is estimated on the basis of a measurement result $x$ using an estimator function $\widetilde{\omega}_x$.
For the optimal performance the average estimation variance
\begin{equation}
\label{eq:bayes}
\Delta^2\widetilde{\omega} = \int \t{d} \omega \t{d} x \, p_\omega \Tr(\rho_\omega \Pi_x) (\omega - \widetilde{\omega}_x)^2
\end{equation}
should be minimized over $\rho$, $\{\Pi_x\}$, $\tilde{\omega}_x$, as well as, the interrogation time $t$.
It was shown \cite{Personick1971} (see also \cite[Chapter VIII]{Helstrom1976}) that the optimal
measurement may be restricted to the class of standard projection von-Neumann measurements $\Pi_x = \ket{x}\bra{x}$,
$\braket{x}{x^\prime} = \delta_{x,x^\prime}$ and the full information on the measurement-estimation strategy is contained in a single
observable $L = \int \tilde{\omega}_x \ket{x}\bra{x} \mathrm{d} x$.
Optimization of $L$ yields the minimal variance
 \begin{equation}
\label{eq:optbayes}
\Delta^2\widetilde{\omega} = \Delta^2\omega - \Tr\left(\bar{\rho} L^2  \right), \, \frac{1}{2}\{L,\bar{\rho}\} = \bar{\rho}^\prime,
\end{equation}
where $\Delta^2\omega$ is the variance of the prior distribution $p_{\omega}$ and
the optimal $L$ is implicitly given by Eq.~\eref{eq:optbayes}, where  $\bar{\rho} = \int \t{d} \omega \, p_\omega \rho_\omega$,
$\bar{\rho}^\prime = \int \t{d}\omega\, p_\omega \rho_\omega \omega$.
A simple derivation of the above formula, together with a new elementary proof of its optimality, is given in \ref{app:optimal}. The optimal estimator $\widetilde{\omega}(x)$ equals the mean of the updated frequency distribution $\propto p_\omega \Tr(\rho_\omega \Pi_x)$, which is the prior for the next experiment.

For a fixed initial state $\rho$ the optimization amounts to solving the anti-commutator Eq.~(\ref{eq:optbayes}) for $L$ and is equivalent in complexity to a computation of QFI for a mixed state. Full optimization requires further optimization of $\Delta^2 \tilde{\omega}$ over $\rho$ and $t$. Interestingly, we have found that an iterative algorithm analogous to the one proposed in \cite{Demkowicz2011} is very effective.
We begin with a random input state,
and iteratively find the optimal measurements and corresponding states. The procedure converges to
optimal solutions and in the cases we have studied outperforms the brute force optimization of the QFI allowing to obtain a solution
in the number of particles regime where the brute-force optimization ceases to be practical on a standard PC ($n \gtrsim 50$). The convergence of the
procedure has been analyzed in \cite{Macieszczak2013}, and its efficiency has been additionally corroborated in
optimization of phase estimation schemes in presence of local dephasing and loss \cite{Jarzyna2014}. See \ref{app:iterative} for details of the implementation of the algorithm.

The similarity between Eqs.~\eref{eq:optbayes} and \eref{eq:cr} becomes even more evident when considering a
gaussian prior distribution $p_\omega \propto \exp(-\omega^2/2 \Delta^2\omega)$. In this case
$\bar{\rho}^\prime = \Delta^2 \omega \int \t{d} \omega p_\omega \frac{\t{d} \rho_\omega}{\t{d} \omega} = \mathrm{i} t \Delta^2 \omega\mathrm  [\bar{\rho}, H]$ and
Eq.~\eref{eq:optbayes} becomes
\begin{equation}
\label{eq:bayesfish}
\Delta^2\widetilde{\omega}(t) = \Delta^2\omega \left[1 - \Delta^2\omega F(\bar{\rho}, H t)\right],
\end{equation}
with $F(\bar{\rho},Ht)$ defined  in Eq.~\eref{eq:fisherH}.
Looking for the optimal probe states in the Bayesian approach is thus equivalent to maximizing QFI for a
spread out state $\bar{\rho}$.

\section{Decoherence-free frequency estimation}
\begin{figure}[t]
\begin{flushright}
\includegraphics[width=0.8\columnwidth]{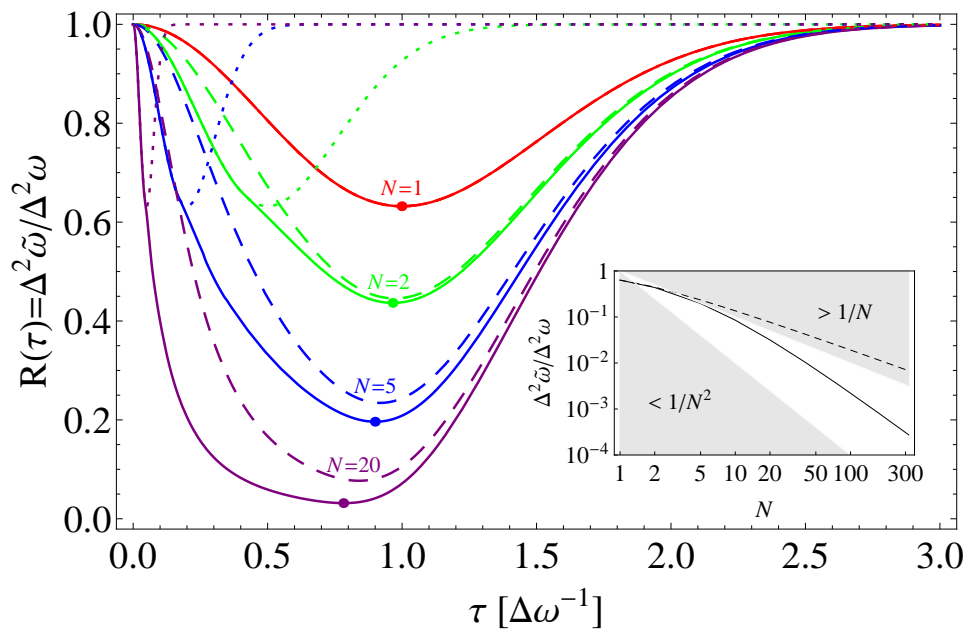}
\caption{Frequency variance reduction for the optimal estimation strategies (solid) compared with
optimized measurement strategies for uncorrelated states (dashed) and the GHZ states (dotted) as a function of evolution
time for a different number of atoms $N$.
Time $\tau$ is given in natural units of inverse prior frequency uncertainty. The inset depicts the dependance of estimation
variance at the optimal time (dotted lines in main
figure) as a function of $N$ for optimal (solid) and uncorrelated (dashed) states.  The shaded areas serve as a guide to the eye to judge the
character of the scaling of precision with $N$. Prior distribution is assumed to be gaussian.}
\label{fig:optimal}
\end{flushright}
\end{figure}

Let us first consider an idealized estimation model, in which $N$ two-level atoms
are subject to unitary evolution $e^{-\mathrm{i} H \omega t}$ with the generator
$H = \sum_{i=1}^N \ket{1}\bra{1}^{(i)}$, where $\ket{1}\bra{1}^{(i)}$ is the projector on the excited state of the $i$-th particle.
Without loss of optimality we may assume that the input probe state is pure $\rho=\ket{\psi}\bra{\psi}$ and
is supported on the  symmetric subspace \cite{Buzek1999}. Let $\ket{n}$ denote a symmetric state with $n$ atoms in $\ket{1}$, then
$\ket{\psi} = \sum_{n=0}^N c_n \ket{n}$ and $H = \sum_n n \ket{n}\bra{n}$.
For a Gaussian prior the averaged state $\bar{\rho}$ in the $\ket{n}$ basis reads:
\begin{equation}
 \bar{\rho}_{nm} =  \rho_{nm}e^{-(n-m)^2 t^2 \Delta^2 \omega/2},
 \end{equation}
where $\rho_{nm}=\bra{n}\rho\ket{m}=c_n c_m^*$.
The results of the numerical minimization of $\Delta^2\tilde{\omega}$ 
 with respect to $\ket{\psi}$
 as a function of interrogation time for different atom numbers are presented in Fig.~\ref{fig:optimal},
where we introduced the optimal variance reduction factor as $R(\tau) = \frac{\Delta^2 \tilde{\omega}(\tau/\sqrt{\Delta^2 \omega})}{\Delta^2 \omega}$
with $\tau$ being the dimensionless time parameter.
Minimal estimation variances achievable with non-entangled
states are presented for comparison. 
The variances which correspond to the optimal interrogation times are depicted in the inset as a function of $N$. For the optimal states the
curve slowly approaches the $1/N$ Heisenberg scaling, while for product states it is limited by $1/\sqrt{N}$.

For small times, $\tau \ll 1/N$, the GHZ states $\ket{\psi}=(\ket{0}+\ket{N})/\sqrt{2}$ minimize $R(\tau)$,
as the effective prior phase distribution is narrow enough not to suffer from the characteristic GHZ $2\pi/N$ ambiguity.
In this regime the exact formula for the minimal variance reads $R(\tau)=1-N^2 \tau^2 \exp(-N^2 \tau^2)$.
For times $\tau \gg 1/N$ 
the optimal states closely approach the ``Sine'' states introduced in \cite{Summy1990, Berry2000},
while in the intermediate regime  the optimal states have a structure which interpolates between the
GHZ and ``Sine'' states. Note also that there is no point in increasing the interrogation time too much, $\tau \gtrsim 1 $, 
as inferring a frequency value is becomes ambiguous and thus the overall estimation variance increases.
Thanks to Eq.~(\ref{eq:bayesfish}) relating the Bayesian cost and the QFI,
one can get a detailed insight into the structure of the optimal states by invoking the results presented in \cite{Knysh2014} where the C-R bound approach in presence of global dephasing has been pursued (see e.g. Fig.~2(i) of \cite{Knysh2014}). One just needs to identify the collective dephasing parameter $\Gamma_0$ from \cite{Knysh2014} with $\tau^2$ in our formulas.
We should also add, that instead of using the exact optimal states, which may be difficult to deal with in practice,
we have checked numerically that the optimal performance may be closely approached using appropriately prepared
one- or two-axis spin squeezed states even though they structure differs significantly from that of the optimal states.s



\section{Frequency estimation under dephasing}
We now consider a model with time dependent $\omega$ which reflects the relevant aspects of atomic clock operation as depicted in Fig.~\ref{fig:scheme}. The
dominant decoherence  effects is the LO noise \cite{Andre2004, Hinkley2013} and the collective dephasing of atoms \cite{Roos2006, Monz2011, Dorner2012}.
Local dephasing of atoms \cite{Huelga1997} or loss \cite{Gross2010} might also be relevant in certain atomic interferometry setups, but we neglect them here for the clarity of presentation.
Let $\omega_{\t{LO}}(t)$ be a stochastic process describing LO noise and $\omega(t) = \omega_{\t{LO}}(t) - \omega_0$ the corresponding detuning. $\omega_{\t{LO}}(0)$ represents the LO frequency which stochastically drifted and was corrected during preceding feedback cycles.
For a given realization of $\omega(t)$ the output state of the atoms after the interrogation time $t$ is $\rho_t = U_t \rho U^\dagger_t$, where $U_t = e^{-\mathrm{i} H \int_0^t \mathrm{d} s (\omega(s) + \Omega(s))}$, and $\Omega(t)$ is the stochastic process representing the
non-LO sources of the collective dephasing of the atoms. Measurement of $\rho_t$ yields a result $x$ with probability equal to $\t{Tr}(\rho_t \Pi_x)$.
The estimator $\tilde{\omega}(x)$ is used to correct the LO frequency. The outcome frequency $\omega(t) - \tilde \omega(x)$ has a variance
%
%
\begin{equation}
\label{eq:variancedecoh}
\Delta^2 \tilde{\omega}(t) = \langle \int \t{d}x \, \t{Tr}(\rho_t\Pi_x) [\omega(t) - \tilde{\omega}(x)]^2 \rangle,
\end{equation}
where $\langle \cdot \rangle$ denotes averaging with respect to the stochastic processes $\omega(t)$ and $\Omega(t)$. Averaging over $\omega(t)$  corresponds to the average over the prior in Eq.~(\ref{eq:bayes}).
The optimal estimation strategy yields the same formula for the minimal variance as in the decoherence-free case above,
 Eq.~\eref{eq:optbayes}, but now with $\Delta^2\omega(t) = \langle \omega(t)^2 \rangle$, $\bar{\rho} = \langle \rho_t \rangle $
and $\bar{\rho}^\prime = \langle \omega(t) \rho_t\rangle$.

When $\Omega(t)$, $\omega(t)$ are independent Gaussian processes with zero means, 
we obtain (see \ref{app:noise} for the derivation):
\begin{equation}
\label{eq:rhobardecoh}
\bar\rho_{nm} = \rho_{nm} e^{-(n-m)^2 t^2 [K_2^\omega(t) + K_2^\Omega(t)]/2},\, \bar \rho' = \mathrm{i}t [\bar \rho, H] K^\omega_1(t),
\end{equation}
where $K^X_2(t) = \frac{1}{t^2} \int_0^t \int_0^t \mathrm{d}t_1 \mathrm{d} t_2 K^X(t_1,\,t_2)$, $K_1^X(t) = \frac{1}{t} \int_0^t \mathrm{d} t_1 K^X(t_1,\,t)$ and $K^X(t_1,\,t_2 )= \langle X(t_1) X(t_2)\rangle$ is the two point correlation function of the process $X(t)$. Again, making a connection to QFI
we have
\begin{equation}
\label{eq:precdecoh}
\Delta^2 \tilde{\omega}(t) = \Delta^2 \omega(t) - K^\omega_1(t)^2 F(\bar \rho, Ht).
\end{equation}
Calculation of $\bar{\rho}$ can be viewed as averaging $\rho$ with respect to a gaussian ``effective prior''
with variance
\begin{equation}
\label{eq:effectiveprior}
\Delta^2 \omega_K(t) = K_2^{\omega}(t) +K^{\Omega}_2(t).
\end{equation}
It is clear that the optimal input state for Bayesian frequency estimation in the presence of decoherence is that which maximizes QFI after
being spread out with the ``effective prior''.
This implies that the solution
\emph{in the presence of decoherence} is in our case already contained in the solution of the \emph{decoherence-free} case and therefore requires no additional numerical
optimization.
Using the variance reduction factor $R(\tau)$ introduced for the decoherence-free case and depicted in Fig.~\ref{fig:optimal},
the formula for the minimal $\Delta^2\tilde{\omega}(t)$ in the presence of decoherence   reads:
\begin{equation}
\label{eq:generaldelta}
\Delta^2 \widetilde{\omega}(t) =
\Delta^2 \omega(t) - \frac{K^\omega_1(t)^2}{\Delta^2 \omega_K(t)}[1- R(t \sqrt{\Delta^2 \omega_K(t)})].
\end{equation}
\begin{figure}[t]
\begin{flushright}
\includegraphics[width=0.8\columnwidth]{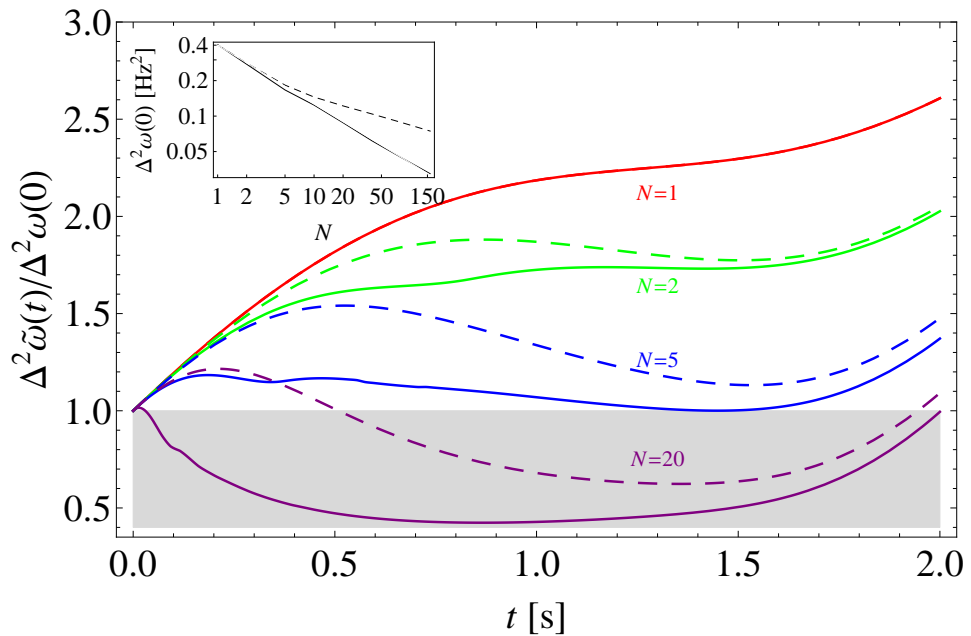}
\caption{Frequency variance reduction for initial variance $\Delta^2\omega(0)=0.167\, \t{Hz}^2$ as a function of interrogation time
for optimal (solid) and product (dashed) probe states for a different number of atoms $N$
 taking into account LO and atomic frequency noise with noise parameters $\alpha=1\, \t{Hz}^2$, $\gamma = 0.2 \,\t{Hz}$, $\beta=0.001 \t{Hz}$.
Strategies in the gray region permit keeping the variance below $\Delta^2\omega(0)$ level. The inset depicts
a log-log plot of the minimal stationary variances $\Delta^2\omega(0)$ as a function of $N$ for the optimal (solid) and product (dashed) probe states.
}
\label{fig:plotdecoh}
\end{flushright}
\end{figure}

As an example consider $\omega(t)=(\omega_{\t{LO}}(0) - \omega_0)e^{-\gamma t} + f^{\t{OU}}(t)$.
This is the Ornstein-Uhlenbeck (OU) process with the initial value $\omega_{\t{LO}}(0)-\omega_0$
and the correlation function of the zero-mean stochastic term
$f^{\t{OU}}(t)$ given by $K^{\t{OU}}(t_1,t_2)=\alpha
 [e^{-\gamma |t_1-t_2|} - e^{-\gamma(t_1+t_2)}]$. We assume that $\omega_{\t{LO}}(0)$ is normally distributed with
 $\langle \omega_{\t{LO}}(0)\rangle=\omega_0$ and is independent from $f^{\t{OU}}(t)$, hence:
$\Delta^2\omega(t) = K^\omega(t,t)=\Delta^2\omega(0)e^{-2 \gamma t} + \alpha(1-e^{-2 \gamma t})$,
which for times smaller than the OU correlation time, $t \ll \gamma^{-1}$,  yields the diffusive character of
frequency distribution broadening $\Delta^2\omega(t) \approx
 \Delta^2 \omega(0) + 2 \gamma t [\alpha -\Delta^2 \omega(0)] $.
The OU process is not perfect in representing noise in real LOs such as lasers, where
$1/f$ noise for low frequencies and white noise for larger frequencies is typical
\cite{Numata2004, Andre2004, Rosenband2013, Jiang2011}. It
may, however, well approximate the real noise in the low frequency regime, with simple analytic formulas for the relevant quantities:
$K^\omega_1(t)=[\Delta^2 \omega(0)(1-e^{-\gamma t})+\alpha(1-e^{-\gamma t})^2]/(\gamma t)$,
$K^\omega_2(t)=[(\Delta^2\omega(0) -\alpha)(e^{-\gamma t}-1)^2+ 2 \alpha(\gamma t +e^{-\gamma t} -1) ]/(\gamma t)^2$.
In order to take into account the white noise contribution present in real systems we
represent it in the atomic dephasing process $\Omega$ with
$K^{\Omega}(t_1,t_2)=\beta \delta(t_1-t_2)$, $K_2^{\Omega}(t) = \beta/t$.
We exclude the white noise part from the LO noise, as this would yield infinite variances $\Delta^2\omega(t)$,
but take it into account as a factor in atomic decoherence.
 This is consistent with the fact that high frequency white noise contributes
to the tails of atomic Lorenzian line shapes without significantly affecting their full-width at half maximum \cite{Domenico2010}.
Using variance as a measure of the width of the line is justified provided only low frequency noise is taken into account.

Treating \cite{Jiang2011} as a guideline we choose 
$\alpha=1\, \t{Hz}^2$, $\gamma = 0.2\, \t{Hz}$ so that
the OU power spectrum approximates reality for small frequencies $\lesssim 10\t{Hz}$ and set $\beta=10^{-3} \t{Hz}$ to represent the effects of high frequency white noise. The
reduction of frequency variance as a function of the interrogation time and the number of atoms used is presented in Fig.~\ref{fig:plotdecoh}.
We took initial frequency variance $\Delta^2 \omega(0)=0.167\, \t{Hz}^{2}$ so that the blue solid curve
representing the optimal quantum strategy for $N=5$
touches the $\Delta^2\omega(t)/\Delta^2\omega(0)=1$ line.
This indicates that a \emph{stationary} operation, where the spread due to frequency noise is
compensated by estimation feedback, is possible with optimally entangled states of $N=5$ atoms. It is also clear that
using product states with the same number of atoms
(blue dashed) or a smaller number of entangled atoms does not permit keeping frequency uncertainty at this level.
For a given $N$  varying $t$ to minimize $\Delta^2 \omega(0)$
while keeping the stationary condition fulfilled yields the optimal stationary strategy.
The inset depicts the minimal stationary $\Delta^2 \omega(0)$ as a function of the number of atoms for optimal (solid) and product states (dashed).
The clear advantage of quantum over classical strategies is evident.
We should also mention that formula
\eref{eq:effectiveprior} allows one to quantify the significance of the prior knowledge in comparison to decoherence effects. Specifically, when
$\Delta^2 \omega(0) \ll \Delta^2\omega_K(t)$, the prior knowledge is not important and the optimal states are determined solely by the effects of decoherence and
will be the same as in the standard QFI approaches  \cite{Dorner2012, Chin2012, Szankowski2012}.

\section{Summary}
To summarize, we have presented a consistent Bayesian approach to studying the potential quantum enhancement in frequency calibration.
The tools presented are general enough to be able to deal with more sophisticated and promising atomic clock setups
including many simultaneously evolving atomic ensembles \cite{Borregaard2013, Rosenband2013}.
However, a rigorous connection between our results and the performance of the actual atomic clock, requires the study of the
Allan variance, instead of instantaneous variance, as a figure of merit. The Allan variance takes into account the effects of the correlations between subsequent interrogation steps, i.e. $\omega_{LO}(0)$ and $\omega_{LO}(t)-\omega_{LO}(0)$, which is more adequate in the
quantification of the performance of atomic clocks but at the same time makes the study of the optimal Bayesian strategies much more involved.
We hope to address this problem in the near future. Actually,
some work has already been done recently in this direction, using bounds on precision rather than considering optimal estimation strategies \cite{Fraas2013} or taking a numerical approach based on the semi-definite programming \cite{Mullan2014}.

\section*{Acknowledgments}
We thank Wojciech Wasilewski, Jan Ko\l{}ody\'nski, Marcin Jarzyna, Andrew Ludlow and M\u{a}d\u{a}lin Gu\c{t}\u{a} for fruitful discussions.
This research was supported by the Polish NCBiR under the ERA-NET CHIST-ERA project QUASAR,
Foundation for Polish Science TEAM project and the FP7 IP project SIQS co-financed by the
Polish Ministry of Science and Higher Education.

\appendix

\section{Optimal quantum Bayesian estimation strategy for the quadratic cost function}
\label{app:optimal}
Here we prove that minimization of the average estimation variance $\Delta^2\widetilde{\omega}$, as given in Eq.~\eref{eq:bayes},
with respect to the choice of measurements $\Pi_x$ and estimators $\widetilde{\omega}_x$ yields Eq.~\eref{eq:optbayes}.
Eq.~\eref{eq:bayes} can be rewritten as:
\begin{equation}
\label{eq:bayesproof}
\Delta^2\widetilde{\omega}=\Delta^2 \omega  + \Tr(\bar{\rho} L_2) - 2 \Tr(\bar{\rho}^{\prime} L),
\end{equation}
where  $\bar{\rho} = \int \t{d} \omega \, p_\omega \rho_\omega$, $\bar{\rho}^\prime = \int \t{d}\omega\, p_\omega \rho_\omega \omega$,
$L = \int \t{d} x \widetilde{\omega}_x\Pi_x $, $L_2 = \int \t{d} x \widetilde{\omega}_x^2\Pi_x $.
First we prove that, without loss of optimality, we can restrict the class of measurements to standard projective von-Neumann measurements, so that
$x \in \{1, \dots, d\}$ where $d$ is the dimension of the relevant Hilbert space
and $\Pi_x = \ket{x}\bra{x}$, $\braket{x}{x^\prime} = \delta_{x,x^\prime}$.

Let $\Pi_x$, $\widetilde{\omega}_x$ be a measurement (not necessarily projective) and the accompanying estimator.
We write the eigendecomposition of the corresponding $L= \int \t{d} x \widetilde{\omega}_x \Pi_x$ as:
\begin{equation}
L = \int \widetilde{\upsilon}_y  \ket{y}\bra{y} \mathrm{d} y = L^{\t{proj}},
\end{equation}
where the eigenprojectors $\ket{y}\bra{y}$ and the eigenvalues $\widetilde{\upsilon}_y$ can now be interpreted as new measurement operators and
estimators respectively; we will
refer to this strategy as the projective strategy.
Below, we will prove that replacing the original strategy with the projective one can only reduce the estimation variance.
Inspecting Eq.~\eref{eq:bayesproof} we notice that information on measurement and estimation strategy enters the formula only through
$L$ and $L_2$ operators. Had the projective strategy been used, we would have obtained:
\begin{equation}
L_2^{\t{proj}} = \int \widetilde{\upsilon}_y^2 \ket{y}\bra{y} \mathrm{d} y = L^2.
\end{equation}
In order to show that we can only benefit by replacing the original strategy with the projective one, it is sufficient to prove that
$L_2 - L^{2} \geq 0$ as this implies $\Tr(\rho L_2) \geq \Tr(\rho L^2)$ and,
consequently, the resulting estimation variance of the projective strategy is not greater than the original one.
More explicitly, we need to demonstrate that:
\begin{equation}
\int \t{d} x \Pi_x \widetilde{\omega}^{2}_x  -  \left( \int \t{d} x \Pi_x \widetilde{\omega}_x \right)^2 \geq 0.
\end{equation}
The above inequality, however, is a special case of operator generalization of Jensen's inequality \cite{Hansen2003} when applied
to the operator convex function $f(t)=t^2$. The operators $\Pi_x$ now play the role of weights in the convex combination.

Having proved the optimality of the projective strategy we can substitute $L_2=L^2$ in Eq.~\eref{eq:bayesproof}
and, as result, face a simple problem of the minimization of:
\begin{equation}
\Delta^2\widetilde{\omega}=\Delta^2 \omega  + \Tr(\bar{\rho} L^2  - 2 \bar{\rho}^{\prime} L)
\end{equation}
with respect to a single hermitian matrix $L$ and no additional constraints on optimization.
Explicit differentiation with regard to matrix elements $L_{ij}$ leads to:
\begin{equation}
\sum_k \bar{\rho}_{ki} L_{jk} + \bar{\rho}_{jk} L_{ki} - 2 \bar{\rho}^\prime_{ji} = 0,
\end{equation}
which in compact notation results in Eq.~\eref{eq:optbayes}.

We note that the optimal estimator $L$ can have both continues or discrete spectra.
The former is the case for example for Gaussian states, where mean position is the parameter to be estimated
in which case $L$ is proportional to the position operator. We should also remark that since we have put no restriction on the allowed measurements, they could be the most general quantum measurements (POVMs) \cite{Holevo1982}, this in particular  includes the adaptive measurement strategies where measurements on part of the atoms depend on the results obtained from previous measurements on the the other part of the atoms, see. e.g. discussion in \cite{Kolodynski2010}. This implies that such adaptive strategies are not advantageous for the present problem as the simple von-Neumann measurement
has been proven optimal.

\section{Efficient iterative procedure for finding the optimal strategy numerically}
\label{app:iterative}
Using Eq.~\eref{eq:optbayes} allows one to find the optimal estimation strategy for a given    input probe state $\rho$ and the corresponding evolved state $\rho_\omega=\Lambda_{\omega}(\rho)$. Finding the optimal strategy, however, also requires the optimization of the input probe state.
The brute force optimization of Eq.~\eref{eq:optbayes} over $\rho$ proves to be highly inefficient.
Here we propose a heuristic iterative procedure that has proved to be extremely efficient in searching for the optimal strategy.
We know that for a particular input probe state the corresponding optimal measurement and estimation strategy is given by the explicit formula in
Eq.~\eref{eq:optbayes}. The opposite is also true. Given a particular measurement and estimator the corresponding optimal
input probe state can easily be found as is demonstrated below.

We start by rewriting Eq.~\eref{eq:bayesproof} as:
\begin{equation}
\Delta^2 \widetilde{\omega} = \Delta^2 \omega  + \Tr\left[\int \t{d} \omega p_\omega \Lambda_\omega (\rho)  (L_2 - 2  \omega L) \right].
\end{equation}
Let us define a map $\Lambda^*_\omega$ that is dual to $\Lambda_\omega$ i.e. $\Tr[\Lambda_\omega(\rho) A] = \Tr[\rho \Lambda^*_\omega(A)]$.
Thus, an equivalent expression for $\Delta^2\widetilde{\omega}$ reads:
\begin{equation}
\Delta^2 \widetilde{\omega} = \Delta^2 \omega + \Tr \left[\rho \int \t{d} \omega p_\omega \Lambda^*_\omega (L_2 - 2  \omega L)  \right].
\end{equation}
Therefore, the optimal input probe state is pure $\rho=\ket{\psi}\bra{\psi}$, where $\ket{\psi}$ should be chosen as the eigenvector
of $\int \t{d} \omega p_\omega \Lambda^*_\omega (L_2 - 2  \omega L)$ operator corresponding to its most negative eigenvalue.

The heuristic procedure we advocate is now as follows. We start with a random input pure state $\rho^{(0)} = \ket{\psi^{(0)}}\bra{\psi^{(0)}}$
(or a state that one expects to be close to the optimal one, but preferably with some small random noise).
Using Eq.$~\eref{eq:optbayes}$ we find the corresponding optimal projective measurement strategy $L^{(0)}$.
We now calculate operator $\int \t{d} \omega p_\omega \Lambda^*_\omega (L^{(0)2} - 2  \omega L^{(0)})$
(where $L^{(0)}_2$ was replaced by $L^{(0)2}$ since the measurement is projective), and find the eigenvector corresponding to its most negative eigenvalue.
This way we obtain $\ket{\psi}^{(1)}$. We repeat the procedure until we arrive at a satisfactory convergence. The results presented in
Fig.~\ref{fig:optimal} were obtained exactly in this way for $\Lambda_\omega(\rho)= e^{-\mathrm{i} H \omega t} \rho e^{\mathrm{i} H \omega t}$
and a gaussian prior $p_\omega$. This method can be applied to any problem involving optimization of the Fisher information with respect to input probe states and is described in detail in \cite{Macieszczak2013}, where also the analysis of the convergence of the algorithm is given.

\section{Minimal frequency estimation variance under LO fluctuations and atomic dephasing---Eqs.(\ref{eq:rhobardecoh},\ref{eq:precdecoh})}
\label{app:noise}
If we rewrite Eq.~\eref{eq:variancedecoh} analogously as in Eq.~\eref{eq:bayesproof}, the LO frequency variance immediately after the
estimation procedure reads:
\begin{equation}
\Delta^2\tilde{\omega}(t) = \langle \omega(t)^2 \rangle + \t{Tr}(\bar{\rho} L_2) -2 \t{Tr}(L \bar{\rho}^\prime),
\end{equation}
where $\bar{\rho} = \langle \rho_t \rangle$, $\bar{\rho}^\prime = \langle \omega(t) \rho_t \rangle$,
 $L = \int \t{d} x \widetilde{\omega}_x\Pi_x $, $L_2 = \int \t{d} x \widetilde{\omega}_x^2\Pi_x $.
Optimization of the estimation strategy proceeds exactly as in the decoherence-free case
and yields the minimal variance:
\begin{equation}
\label{eq:optdecohder}
\Delta^2\tilde{\omega}(t) = \langle \omega(t)^2 \rangle - \t{Tr}(\bar{\rho} L^2), \ \frac{1}{2}\{L, \bar{\rho}\}= \bar{\rho}^\prime.
\end{equation}
Matrix elements of $\bar{\rho}$ read:
\begin{equation}
\bar{\rho}_{nm} = \rho_{nm} \langle e^{-\mathrm{i}(n -m)  \int_{0}^t \t{d}s\, ( \omega(s)+\Omega(s))} \rangle.
\end{equation}
Since $\omega(t)$ and $\Omega(t)$ are independent Gaussian processes with zero means, the standard cummulant expansion method yields:
\begin{equation}
\bar{\rho}_{nm}=\rho_{nm}
e^{-\frac{(n-m)^2}{2}\int_{0}^t \t{d}s_1 \t{d}s_2 \, \langle (\omega(s_1)+\Omega(s_1)( \omega(s_2)+\Omega(s_2))\rangle  }
=\rho_{nm}e^{-\frac{(n-m)^2 t^2}{2} (K_2^{\omega} + K_2^{\Omega})}.
\end{equation}
In order to apply the cummulant expansion method to the calculation of $\bar{\rho}^\prime$, we
write its matrix elements as
\begin{equation}
\bar{\rho}^\prime_{nm}=\rho_{nm} \frac{\t{d}}{\t{d} \xi}
\left.\left\langle e^{\int_{0}^t \t{d}s\,  \mathrm{i}(m -n)[ \omega(s)+\Omega(s)] + \xi \omega(s)\delta(s-t)}\right\rangle \right|_{\xi=0},
\end{equation}
where $\delta(s-t)$ is the Dirac delta. This gives us:
\begin{equation}
\bar{\rho}^{\prime}_{nm} = \bar{\rho}_{nm} \frac{\t{d}}{\t{d} \xi} \left.  e^{-\xi \mathrm{i}(n-m)t K_1^{\omega}(t)- \frac{\xi^2 t^2}{2}K^{\omega}(t,t)} \right|_{\xi=0}
= - \mathrm{i} t \bar{\rho}_{nm}(n-m) K_1^{\omega}(t),
\end{equation}
which also reads as:
\begin{equation}
\bar{\rho}^\prime = \mathrm{i} t [\bar{\rho}, H] K^\omega_1(t).
\end{equation}
Taking into account the definition of the Fisher information, Eqs. (\ref{eq:cr}), (\ref{eq:fisherH}) and Eq.~\eref{eq:optdecohder},
we obtain the desired formula:
\begin{equation}
\Delta^2 \tilde{\omega}(t) = \Delta^2 \omega(t) - K^\omega_1(t)^2 F(\bar \rho, Ht).
\end{equation}

\bibliographystyle{iopart-num}

\begin{thebibliography}{10}
\expandafter\ifx\csname url\endcsname\relax
  \def\url#1{{\tt #1}}\fi
\expandafter\ifx\csname urlprefix\endcsname\relax\def\urlprefix{URL }\fi
\providecommand{\eprint}[2][]{\url{#2}}

\bibitem{Kleppner2006}
Kleppner D 2006 {\em Physics Today\/} {\bf 59} 10

\bibitem{Chou2010}
Chou C~W, Hume D~B, Rosenband T and Wineland D~J 2010 {\em Science\/} {\bf 329}
  1630--1633

\bibitem{Hinkley2013}
Hinkley N, Sherman J~A, Phillips N~B, Schioppo M, Lemke N~D, Beloy K, Pizzocaro
  M, Oates C~W and Ludlow A~D 2013 {\em Science\/} {\bf 341} 1215--1218

\bibitem{Ramsey1980}
Ramsey N~F 1980 {\em Physics Today\/} {\bf 33} 25--30

\bibitem{Lee2002}
Lee H, Kok P and Dowling J~P 2002 {\em Journal of Modern Optics\/} {\bf 49}
  2325--2338

\bibitem{Wineland1992}
Wineland D~J, Bollinger J~J, Itano W~M, Moore F~L and Heinzen D~J 1992 {\em
  Phys. Rev. A\/} {\bf 46}(11) R6797--R6800

\bibitem{Holland1993}
Holland M~J and Burnett K 1993 {\em Phys. Rev. Lett.\/} {\bf 71} 1355--1358

\bibitem{Bollinger1996}
Bollinger J~J, Itano W~M, Wineland D~J and Heinzen D~J 1996 {\em Phys. Rev.
  A\/} {\bf 54} R4649--R4652

\bibitem{Berry2000}
Berry D~W and Wiseman H~M 2000 {\em Phys. Rev. Lett.\/} {\bf 85} 5098--5101

\bibitem{Giovannetti2006}
Giovannetti V, Lloyd S and Maccone L 2006 {\em Phys. Rev. Lett.\/} {\bf 96}
  010401

\bibitem{Roos2006}
Roos C~F, Chwalla M, Kim K, Riebe M and Blatt R 2006 {\em Nature\/} {\bf 443}
  316--319

\bibitem{LIGO2011}
LIGO{\;}Collaboration 2011 {\em Nature Phys.\/} {\bf 7} 962--965

\bibitem{Ospelkaus2011}
Ospelkaus C, Warring U, Colombe Y, Brown K~R, Amini J~M, Leibfried D and
  Wineland D~J 2011 {\em Nature\/} {\bf 476} 181--184

\bibitem{Sewell2012}
Sewell R~J, Koschorreck M, Napolitano M, Dubost B, Behbood N and Mitchell M~W
  2012 {\em Phys. Rev. Lett.\/} {\bf 109}(25) 253605

\bibitem{Caves1981}
Caves C~M 1981 {\em Phys. Rev. D\/} {\bf 23} 1693--1708

\bibitem{Huver2008}
Huver S~D, Wildfeuer C~F and Dowling J~P 2008 {\em Phys. Rev. A\/} {\bf 78}
  063828

\bibitem{Dorner2009}
Dorner U, Demkowicz-Dobrza{\'n}ski R, Smith B~J, Lundeen J~S, Wasilewski W,
  Banaszek K and Walmsley I~A 2009 {\em Phys. Rev. Lett.\/} {\bf 102} 040403

\bibitem{Demkowicz2009}
Demkowicz-Dobrzanski R, Dorner U, Smith B~J, Lundeen J~S, Wasilewski W,
  Banaszek K and Walmsley I~A 2009 {\em Phys. Rev. A\/} {\bf 80}(1) 013825

\bibitem{Knysh2010}
Knysh S, Smelyanskiy V~N and Durkin G~A 2011 {\em Phys. Rev. A\/} {\bf 83}
  021804

\bibitem{Demkowicz2013}
Demkowicz-Dobrza\'{n}ski R, Banaszek K and Schnabel R 2013 {\em Phys. Rev. A\/}
  {\bf 88}(4) 041802

\bibitem{Huelga1997}
Huelga S~F, Macchiavello C, Pellizzari T, Ekert A~K, Plenio M~B and Cirac J~I
  1997 {\em Phys. Rev. Lett.\/} {\bf 79} 3865--3868

\bibitem{Genoni2011}
Genoni M~G, Olivares S and Paris M~G~A 2011 {\em Phys. Rev. Lett.\/} {\bf
  106}(15) 153603

\bibitem{Dorner2012}
Dorner U 2012 {\em New Journal of Physics\/} {\bf 14} 043011

\bibitem{Shaji2007}
Shaji A and Caves C~M 2007 {\em Phys. Rev. A\/} {\bf 76} 032111

\bibitem{Sarovar2006}
Sarovar M and Milburn G~J 2006 {\em J. Phys. A: Math. Gen.\/} {\bf 39} 8487

\bibitem{Giovannetti2011}
Giovannetti V, Lloyd S and Maccone L 2011 {\em Nature Photon.\/} {\bf 5}
  222--229

\bibitem{Banaszek2009}
Banaszek K, Demkowicz-Dobrza\'{n}ski R and Walmsley I~A 2009 {\em Nature
  Photon.\/} {\bf 3} 673--676

\bibitem{Escher2011}
Escher B~M, de~Matos~Filho R~L and Davidovich L 2011 {\em Nature Phys.\/} {\bf
  7} 406--411

\bibitem{Demkowicz2012}
Demkowicz-Dobrza\'{n}ski R, Ko\l{}ody\'{n}ski J and {Gu{\c t}{\u a}} M 2012
  {\em Nat. Commun.\/} {\bf 3} 1063

\bibitem{Kolodynski2013}
Ko\l{}ody\'{n}ski J and Demkowicz-Dobrza\'{n}ski R 2013 {\em New Journal of
  Physics\/} {\bf 15} 073043

\bibitem{Chin2012}
{Chin} A~W, {Huelga} S~F and {Plenio} M~B 2012 {\em Phys. Rev. Lett.\/} {\bf
  109} 233601

\bibitem{Szankowski2012}
{Szankowski} P, {Chwedenczuk} J and {Trippenbach} M 2012 {\em ArXiv e-prints\/}
  (\textit{Preprint} \eprint{1212.2528})

\bibitem{Buzek1999}
Bu\v{z}ek V, Derka R and Massar S 1999 {\em Phys. Rev. Lett.\/} {\bf 82}(10)
  2207--2210

\bibitem{Andre2004}
Andr{\'e} A, S{\o}rensen A~S and Lukin M~D 2004 {\em Phys. Rev. Lett.\/} {\bf
  92} 230801

\bibitem{Leibfried2004}
Leibfried D, Barrett M~D, Schaetz T, Britton J, Chiaverini J, Itano W~M, Jost
  J~D, Langer C and Wineland D~J 2004 {\em Science\/} {\bf 304} 1476--1478

\bibitem{Mullan2012}
Mullan M and Knill E 2012 {\em Quantum Info. Comput.\/} {\bf 12} 553--574 ISSN
  1533-7146

\bibitem{Borregaard2013b}
Borregaard J and S\o{}rensen A~S 2013 {\em Phys. Rev. Lett.\/} {\bf 111}(9)
  090801

\bibitem{Helstrom1976}
Helstrom C~W 1976 {\em Quantum detection and estimation theory\/} (Academic
  press)

\bibitem{Braunstein1994}
Braunstein S~L and Caves C~M 1994 {\em Phys. Rev. Lett.\/} {\bf 72} 3439--3443

\bibitem{Fraas2013}
{Fraas} M 2013 {\em ArXiv e-prints\/} (\textit{Preprint} \eprint{1303.6083})

\bibitem{Mullan2014}
{Mullan} M and {Knill} E 2014 {\em ArXiv e-prints\/} (\textit{Preprint}
  \eprint{1404.3810})

\bibitem{Holevo1982}
Holevo A~S 1982 {\em Probabilistic and Statistical Aspects of Quantum Theory\/}
  (North Holland, Amsterdam)

\bibitem{Demkowicz2011}
Demkowicz-Dobrza{\'n}ski R 2011 {\em Phys. Rev. A\/} {\bf 83} 061802

\bibitem{Personick1971}
Personick S 1971 {\em Information Theory, IEEE Transactions on\/} {\bf 17}
  240--246 ISSN 0018-9448

\bibitem{Macieszczak2013}
{Macieszczak} K 2013 {\em ArXiv e-prints\/} (\textit{Preprint}
  \eprint{1312.1356})

\bibitem{Jarzyna2014}
{Jarzyna} M and {Demkowicz-Dobrzanski} R 2014 {\em ArXiv e-prints\/}
  (\textit{Preprint} \eprint{1407.4805})

\bibitem{Summy1990}
Summy G~S and Pegg D~T 1990 {\em Opt. Commun.\/} {\bf 77} 75--79

\bibitem{Knysh2014}
Knysh S~I, Chen E~H and Durkin G~A 2014 {\em ArXiv e-prints\/}  arXiv:1402.0495

\bibitem{Monz2011}
Monz T, Schindler P, Barreiro J~T, Chwalla M, Nigg D, Coish W~A, Harlander M,
  H\"ansel W, Hennrich M and Blatt R 2011 {\em Phys. Rev. Lett.\/} {\bf
  106}(13) 130506

\bibitem{Gross2010}
Gross C, Zibold T, Nicklas E, Esteve J and Oberthaler M~K 2010 {\em Nature\/}
  {\bf 464} 1165--1169

\bibitem{Numata2004}
Numata K, Kemery A and Camp J 2004 {\em Phys. Rev. Lett.\/} {\bf 93}(25) 250602

\bibitem{Rosenband2013}
{Rosenband} T and {Leibrandt} D~R 2013 {\em ArXiv e-prints\/}
  (\textit{Preprint} \eprint{1303.6357})

\bibitem{Jiang2011}
{Jiang} Y~Y, {Ludlow} A~D, {Lemke} N~D, {Fox} R~W, {Sherman} J~A, {Ma} L~S and
  {Oates} C~W 2011 {\em Nature Photonics\/} {\bf 5} 158--161

\bibitem{Domenico2010}
Domenico G~D, Schilt S and Thomann P 2010 {\em Appl. Opt.\/} {\bf 49}
  4801--4807

\bibitem{Borregaard2013}
{Borregaard} J and {S{\o}rensen} A~S 2013 {\em ArXiv e-prints\/}
  (\textit{Preprint} \eprint{1304.5944})

\bibitem{Hansen2003}
Hansen F and Pedersen G~K 2003 {\em Bulletin of the London Mathematical
  Society\/} {\bf 35} 553--564

\bibitem{Kolodynski2010}
Ko\l{}ody\'{n}ski J and Demkowicz-Dobrza\'{n}ski R 2010 {\em Phys. Rev. A\/}
  {\bf 82}(5) 053804

\end{thebibliography}
\providecommand{\newblock}{}

\end{document}